 \documentclass[%
  reprint,
  amsmath,amssymb,
  aps,prl, secnumarabic, graphics,floatfix,nofootinbib,
tightenlines,nobibnotes
 ]{revtex4-1}

\usepackage{graphicx}% Include figure files
\usepackage{dcolumn}% Align table columns on decimal point
\usepackage{bm}% bold math
\usepackage{verbatim}
\usepackage{enumerate}
\usepackage{color}
\usepackage{ulem}
\usepackage[caption=false]{subfig}
\usepackage{todonotes}
\usepackage{SIunits}
\usepackage{notes2bib}
\usepackage[colorlinks=true, urlcolor=blue, linkcolor=blue, citecolor=blue]{hyperref}

\DeclareMathOperator{\sech}{sech}

\begin{document}

\preprint{APS/123-QED}

\title{Non-Hamiltonian Dynamics of Quantized Vortices in Bose-Einstein Condensates}% Force line breaks with \\
%\thanks{}%

\author{Scott A.~Strong  }
\email{sstrong@mines.edu}
\author{Lincoln D.~Carr}
 \affiliation{Department of Physics, Colorado School of Mines}%Lines break automatically or can be forced with \\

%\affiliation{%

\begin{comment}
\author{Charlie Author}
 \homepage{http://www.Second.institution.edu/~Charlie.Author}
\affiliation{
 Second institution and/or address\\
 This line break forced% with \\
}%
\affiliation{
 Third institution, the second for Charlie Author
}%
\author{Delta Author}
\affiliation{%
 Authors' institution and/or address\\
 This line break forced with \textbackslash\textbackslash
}%

\collaboration{CLEO Collaboration}%\noaffiliation
% 
\end{comment}
\date{\today}% It is always \today, today,

\begin{abstract}
The dynamics of quantized vortices in weakly interacting superfluids are often modeled by a nonlinear Schr\"odinger equation. In contrast, we show that quantized vortices in fact obey a non-Hamiltonian evolution equation, which enhances dispersion along the vortex line while introducing a gain mechanism. This allows the vortex medium to support a helical shock front propagating ahead of a dissipative soliton. This dynamic relaxes localized curvature events into Kelvin wave packets. Consequently, a beyond local induction model provides a pathway for decay in low-temperature quantum turbulence.
% \begin{description}	
% Secondary publications and information retrieval purposes.
% \item[PACS numbers]
% May be entered using the \verb+\pacs{#1}+ command.
% You may use the \texttt{description} environment to structure your abstract;
% use the optional argument of the \verb+\item+ command to give the category of each item. 
% \end{description}
\end{abstract}

\pacs{Valid PACS appear here}% PACS, the Physics and Astronomy
\maketitle

%\tableofcontents

Quantized vortices are slender, non-diffusive regions of low density about which the superfluid bulk circulates at strengths defined by multiples of Plank's constant scaled by a characteristic mass~[\onlinecite{Tsubota2013QuantizedTurbulence}]. The three-dimensional Gross-Pitaevskii equation of mean-field theory restricts this density depletion to a formally one-dimensional subregion, of the otherwise irrotational gas. This vortex defect compromises the connectedness of the condensate in accordance with Helmholtz theorem.  If vorticity acts as a primitive source of a fluid flow, then quantum liquids are a setting where a complete theory of the topological hydrodynamics is most likely. Such a theory will resolve mean-field physics at the scale of the vortex core with that of the disordered arrangement of persistent and stable vortex structures with homogeneous circulation characterizing quantum turbulence. This quantum tangle supports various cascade processes which transfer energy between the spatial scales~[\onlinecite{Barenghi2014IntroductionTurbulence.}]. Unique to quantum fluids is the Kelvin wave cascade which relaxes high curvature cusps to small wavelength helical excitations along the vortex. These waves transport turbulent energy to the boundaries~[\onlinecite{Leadbeater2001SoundReconnections}]. In this Letter, we derive a relatively simple, but non-Hamiltonian evolution of the geometric properties of a vortex defect under  \textit{local induction models}. This evolution approximates dynamics at intermediate scales and predicts that the vortex medium transforms an initially localized curvature soliton into a helical shock wave where a dissipative soliton travels behind helical excitations. Gain mechanisms introduced by the non-Hamiltonian structure results in an increase of average curvature and signifies the emergence of the small-scale structures necessary for acoustic emission. 

\begin{figure}
\includegraphics[width=0.5\textwidth]{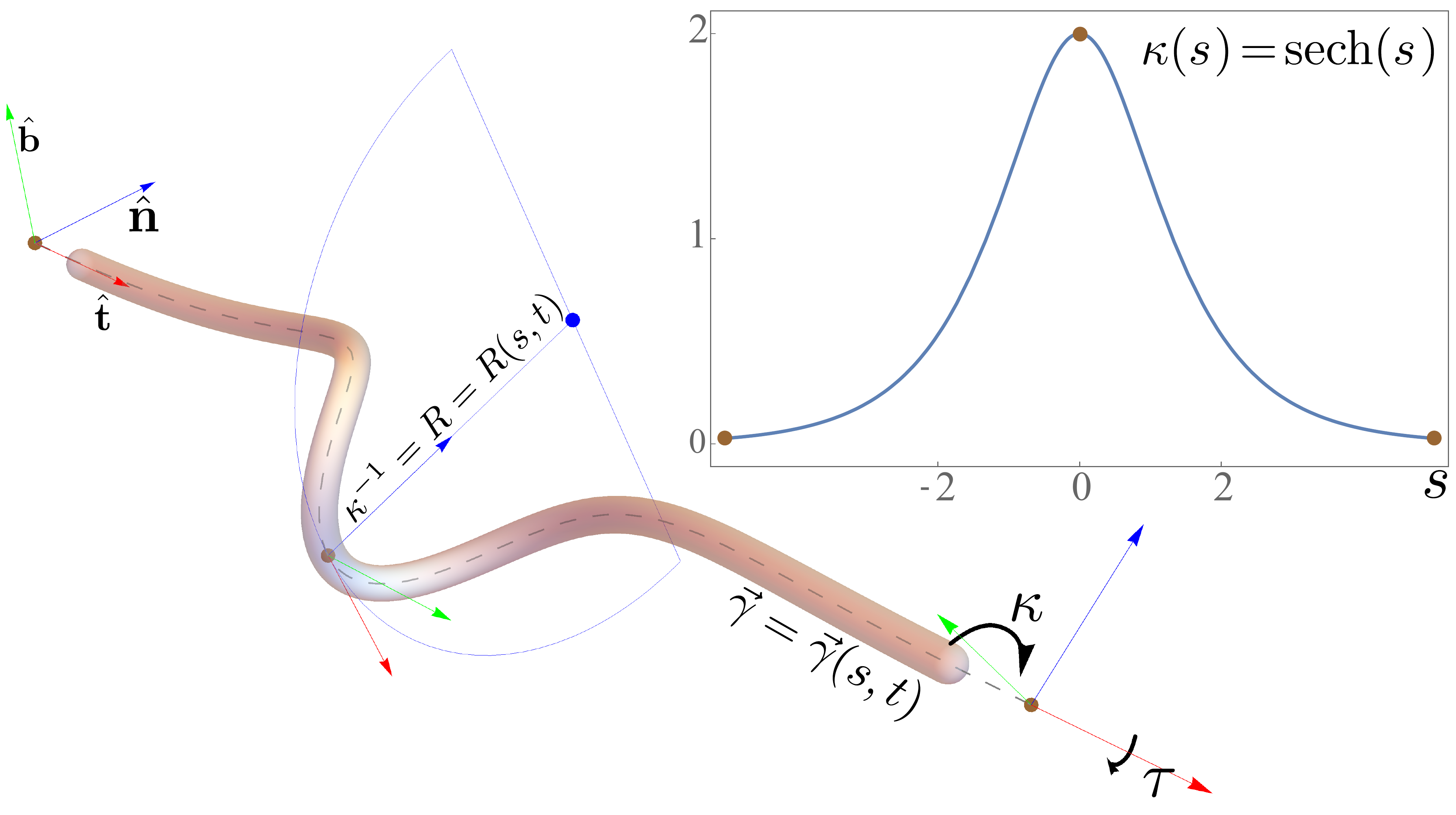}
\caption{Hasimoto's Vortex Soliton. We depict the Frenet frame, $\hat{\textbf{t}}, \hat{\textbf{n}}, \hat{\textbf{b}}$, for a hyperbolic secant (inset) bright soliton, $\kappa, \tau=1$, from the integrable theory, $2i\psi_{t} + 2\psi_{ss}+|\psi|^{2}\psi=0$, corresponding to Hasimoto's map applied to the local induction approximation, $\vec{\gamma}_{t} = \kappa \hat{\textbf{b}}$.
}
\label{fig:schematics}
\end{figure}

The Biot-Savart integral provides a representation of the velocity field by ``un-curling'' the vorticity field. Locally induced evolutions are given by asymptotic approximations to regularizations of this singular integral. There are several regularization techniques available and all yield a lowest order  \textit{local induction approximation}, which states that the binormal component of the velocity field is proportional to local curvature~[\onlinecite{Ricca1996TheDynamics}]. Applying Hasimoto's transform to the local induction approximation defines a cubic focusing nonlinear Schr\"odinger evolution of the curvature and torsion of the vortex~[\onlinecite{Hasimoto1972}].  This theory predicts a bright curvature soliton defining a traveling kink on the vortex line. We call the vortex state, corresponding to this integrable dynamic of the geometric properties, a  \textit{Hasimoto vortex soliton}, {see Fig.~(\ref{fig:schematics}).} We study breakdown in the integrable theory through a non-Hamiltonian description which introduces a curvature gain/loss mechanism while enhancing dispersion on the vortex medium. This model evolves the Hasimoto vortex soliton towards a log-normal type distribution, which we call a \textit{cascade soliton}. Represented on the vortex line, the soliton kink decreases its amplitude while the curvature of the straight background increases ahead of the disturbance. Our non-Hamiltonian dynamic predicts the existence of two qualitatively different dynamical regimes, which are confirmed via simulation. For small perturbations, the kink maintains its structure for longer times, a robustness that is indicative of a dissipative soliton.  Increasing dispersion erodes the previously robust kink into a packet of Kelvin waves generating a profile similar to a dispersive shock wave. This Letter introduces a fully nonlinear integro-differential equation and uses it to approximate the non-Hamiltonian dynamics about the bright soliton fixed point of the integrable theory. Prior to simulations, we consider the dynamics predicted by the emergent gain/loss mechanism in conjunction with the changes to plane wave dispersion to understand the short-time behavior of a soliton.  

{Vortex filament methods} simulate a quantum fluid by evolving its vortical skeleton according to the Biot-Savart integral and are significantly more efficient than mean-field methods for vortex dominated flows~[\onlinecite{Tsubota2017NumericalTurbulence}]. Recent simulations by Salman demonstrate that evolutions given by the mean-field, Biot-Savart and induction models are consistent up to the point where the smallest length scales dominate the physics, a known limitation of filament methods~[\onlinecite{Salman2013BreathersVortices}]. At the same time, Bustamante and Nazarenko have shown that the Biot-Savart integral manifests from Gross-Pitaevskii mean--field dynamics and provides a self-consistent regularization procedure~[\onlinecite{Bustamante2015DerivationEquation}]. This allows us to derive a non-Hamiltonian evolution consistent with a locally induced flow generated by a region of vortex whose arc-length is on the order of the condensate healing length. Our prediction of helical waves generated from a localized curvature event is consistent with current models of energy transfer in the highly quantum turbulent regime~[\onlinecite{Walmsley2014DynamicsSpectra}].   

{As shown in Fig.~(\ref{fig:schematics}),} we define $\vec{\gamma}\in\mathbb{R}^{3}$ as the set of points in three-space {corresponding to a Hasimoto vortex soliton, i.e., a bright curvature soliton shown in the inset of Fig.~(\ref{fig:schematics}).}  The {vortex is} parameterized by an arc-length, $s$, and changes with time, $t$, so that $\vec{\gamma}=\vec{\gamma}(s,t)$. In our previous work we derive, from the Biot-Savart integral, an exact representation of the velocity field induced by a plane circular arc of vorticity~[\onlinecite{Strong2012GeneralizedTurbulence}]. The corresponding asymptotic representation, taken to lowest order in curvature, is equivalent to the local induction approximation. Specifically, the non-circulatory and non-axial velocity field neighboring an arbitrary reference point on the vortex line generates a flow satisfying the vector evolution law,
\begin{align}\label{eqn:BNF}
\frac{\partial \vec{\gamma}}{\partial t} = \alpha \left( \frac{\partial \vec{\gamma}}{\partial s} \times \frac{\partial^{2} \vec{\gamma}}{\partial s^{2}}\right),
\end{align}
where the cross-product  $\vec{\gamma}_{s}\times\vec{\gamma}_{ss}=\kappa \hat{\bm{b}}$ is defined using the binormal vector of the Frenet-Serret frame and $\kappa=\kappa(s,t)$ is the local curvature. The Hasimoto vortex soliton is a prediction of the local induction approximation, given by Eq.~(\ref{eqn:BNF}) where $\alpha=1$, and forms a propagating curvature disturbance, $\kappa(s,t) = 2 \sech(s-t)$, with constant torsion, $\tau=1$. To understand the fragility of the integrable model, and its transition to a cascade soliton, we consider the case of non-constant $\alpha$. 

The condensate healing length, $\xi$, defines the vortex core size and its product with the characteristic curvature, $\kappa$, yields the small parameter $\epsilon = \xi \kappa \ll 1$. The healing length, in ratio with the characteristic system size, $d$,  defines the parameter $\delta = d/\xi$ and is large when the characteristic system size is taken to be the condensate width. The proportionality constant in Eq.~(\ref{eqn:BNF}) is a function of the dimensionless parameters, $\alpha = \alpha(\delta, \epsilon)$, and has a tidy representation given by the matched asymptotic expansion~[\onlinecite{Fetter2001VorticesCondensate}],
\begin{align}\label{eqn:SV}
\alpha(\delta,\epsilon) = \frac{\Gamma}{4\pi} \left[\ln\left(\frac{1}{\delta}\right) +\ln\left(\sqrt{1+ \frac{\epsilon^{2} \delta^2}{8}}\right)\right],
\end{align}
where $\Gamma$ measures the strength of the condensate circulation about the vortex line. The local induction approximation retains only the curvature-independent logarithmic singularity and is defined as Eq.~(\ref{eqn:BNF}), where  $\alpha=\alpha(\delta,\epsilon=0)$. If we assume that $\epsilon$ is not formally zero, then the scale separation $\epsilon \ll \epsilon \delta \ll \delta$ defines a regime where the linearization of Eq.~(\ref{eqn:SV}), i.e. the local induction approximation, is accurate. {If a parameterization is given, then the characteristic length, $d$, can be associated with the domain of Biot-Savart integration. Parameterizing an arbitrary vortex element by a plane circular arc gives a wide range for the non-dimensional parameter, $ \delta \in (0.3416293,  100)$. The lower bound of this interval is given by the Bustamante-Nazarenko regularization, while the upper bound corresponds to a vortex ring with a radius on the order of tens of micrometers. While this approach can maintain the previous scale separation, it also permits the study of flows induced by arcs of vorticity with small central angle such that $\epsilon \delta=O(\epsilon)$. We specifically consider motion induced by filament elements whose length is near the new self-consistent cutoff~[\onlinecite{Bustamante2015DerivationEquation}].}

\begin{figure}
 \includegraphics[width=0.5\textwidth]{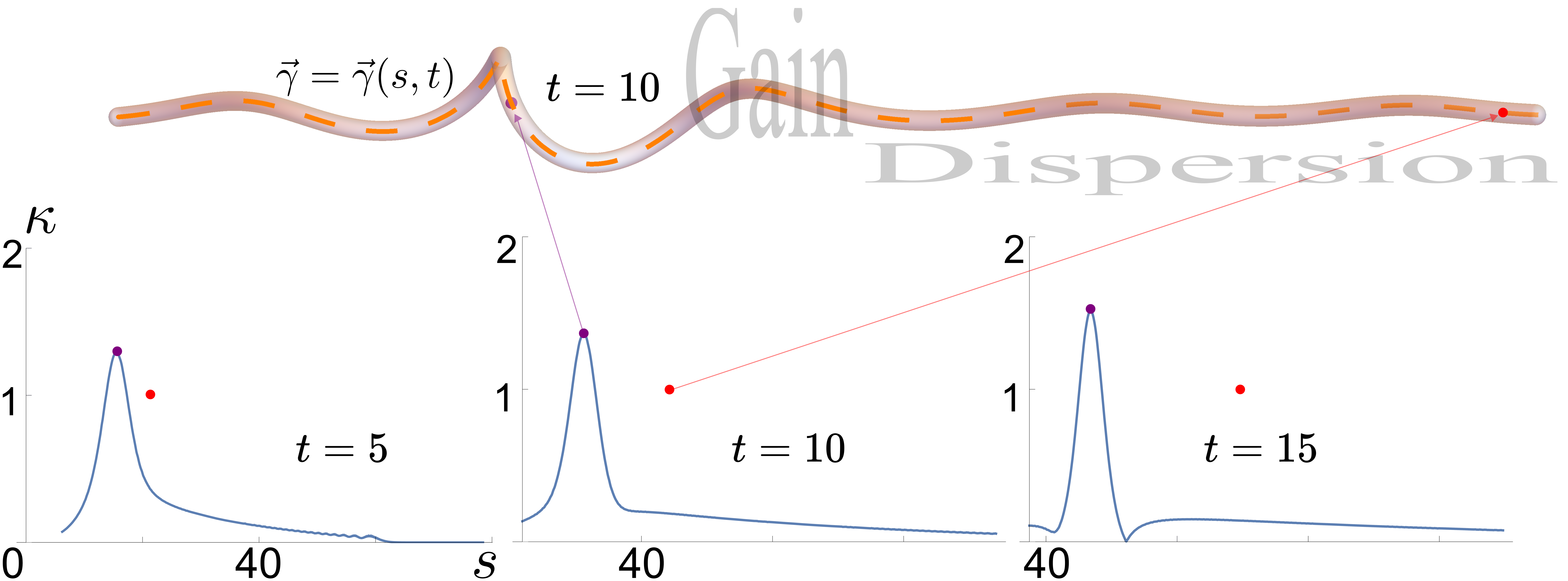}
 \caption{Non-Hamiltonian Cascade Soliton. An initial Hasimoto vortex soliton experiences dispersion producing helical waves propagating ahead of the soliton kink. The non-Hamiltonian gain mechanism supports both the kink and the helical excitations as the localized curvature event is transitioned to a cascade process. 
 }
\label{fig:schematics2}
\end{figure}

Hasimoto's transformation rotates the Frenet basis into $\mathbb{C}^{3}$ where the {geometric wave function}, $ \psi(s,t) = \kappa(s,t)~ \mbox{Exp}\left[{i \int^{s}_{0}ds' \tau(s',t)}\right]$, carries the curvature and torsion variables of a vortex line satisfying local induction approximation~[\onlinecite{Hasimoto1972}].  
This transformation is robust and can be used to map more general flows, often leading to complicated integro-differential equations~[\onlinecite{Fukumoto1991Three-dimensionalVelocity},\onlinecite{Majda2002VorticityFlow}]. Restricting ourselves to binormal flows defined by Eq.~(\ref{eqn:BNF}) and applying the  asymptotic representations of $\alpha$ we recast the corresponding integro-differential equation as a fully nonlinear differential equation. {The predicted dynamical state is a cascade soliton, which is a solitary wave accompanied by helical excitations, and is depicted in Fig. (\ref{fig:schematics2}).}

Measuring time in units defined by the ratio of characteristic vortex with curvature, fluid circulation, and the first term in the expansion of $\alpha$, $\tilde{t}=4\pi\tilde{\gamma}/(\tilde{\kappa} \Gamma a_{0}(\delta))$, gives $\alpha(\delta,0)=1$. Application of the Hasimoto transformation to the general case, $\alpha=\alpha(\delta, \epsilon)$ yields a nonlinear integro-differential evolution of the geometric variables,
\begin{align}\label{eqn:GHNLS}
&i \psi_{t} +  \left[\alpha\psi\right]_{ss} +\frac{\alpha}{2} |\psi|^{2}\psi +\frac{\psi}{2} \int^{s}  |\psi|^{2}   \alpha_{s'}\,  ds'=0.
 \end{align}
  This evolution maintains a Schr\"odinger structure, however, it is difficult to derive useful information from it. The small parameter $\epsilon \ll 1$ provides an expansion of $\alpha$ in powers of $\kappa$. For  Eq.~(\ref{eqn:SV}), we find that $\alpha(-\kappa)=\alpha(\kappa)$. {After truncating quartic and higher terms we arrive at the}  simpler evolution,
\begin{align}\label{eqn:GHNLSPDE1}
&i \psi_{t} + \psi_{ss} + \frac{1}{2}|\psi|^{2} \psi +  \lambda \left(  \left[|\psi|^2 \psi\right]_{ss}+ \frac{3}{4}|\psi|^{4}\psi \right) = 0,
\end{align}
which we call the \textit{non-Hamiltonian vortex cascade equation} (NVC). The {correction parameter}, $\lambda$, depends on our dimensionless constants and is the ratio of the second and first coefficients in the expansion of $\alpha$. In the local induction approximation $\lambda=0$ and we have an integrable theory. If $\lambda \neq 0$, then integrability is compromised so severely that nearly all underlying symmetries are broken. With the exception of arc-length, a quantity conserved by the binormal flow itself, nothing typical, {like energy or momentum}, is conserved. 

To understand this loss of mathematical structure we inspect the the fully nonlinear term $\left[|\psi|^{2} \psi\right]_{ss}$. One can show that there does not exist a functional whose variational derivative satisfies the necessary self-adjoint conditions and therefore the evolution cannot be written as an infinite dimensional Hamiltonian system~[\onlinecite{Olver2000ApplicationsEquations},\onlinecite{Faddeev2007HamiltonianSolitons}]. The system is invariant with respect to arbitrary time and space translations. However as the system is not Hamiltonian, Noether's theorem does not apply and our continuous symmetries need not correspond to conserved densities.  Application of the SYM symmetry software package~[\onlinecite{Dimas2006APDEs}] to Eq.~(\ref{eqn:GHNLSPDE1}) found no additional continuous symmetries. Additionally, a Mathematica package that symbolically calculates conservation laws found no low-order conserved densities~[\onlinecite{Poole2011SymbolicDimensions}]. While the system does possess discrete parity and time symmetries, we consider non-symmetric, time-irreversible dynamics of the nonlinear evolution.

Without $\left[|\psi|^{2} \psi\right]_{ss}$, Eq.~(\ref{eqn:GHNLSPDE1}) is a  complex quintic Ginzburg-Landau equation used in the study of dissipative solitons~[\onlinecite{Akhmediev2005DissipativeSolitons}]. Our real coefficients imply a Hamiltonian structure and a nonlinear gain/loss mechanism must enter through other means. Madelung's transformation decomposes Schr\"odinger evolutions into real and imaginary parts~[\onlinecite{Madelung1926EineSchrodinger}]. {Transforming Eq.~(\ref{eqn:GHNLSPDE1}) yields a system of first-order evolutions on the bending density, $\rho=\kappa^{2}=|\psi|^{2}$ and torsion. From this system we find that the total bending across the vortex,
\begin{align}\label{eqn:L2}
\frac{d}{dt}\int_{a}^{b} \kappa^{2} \, ds &=-2\lambda \int_{a}^{b} \rho_{s} \rho \tau \, ds, 
\end{align}
 is no longer conserved in time. Specifically, the bending dynamics arise from the correction term and are, in part, determined by the helicity density, $\rho \tau$. This density corresponds to the momentum density in the condensate picture, and is also not conserved by the NCV for superpositions of helical modes. This non-Hamiltonian geometric gain/loss mechanism provides a pathway for dissipative soliton dynamics. For example, if the torsion is positive, then bending energy grows/decays over regions where curvature is decreasing/increasing. Simulations indicate that this feature is robust against  distortions manifesting from plane wave dispersion.}  
 
\begin{figure}
  \includegraphics[width=0.5\textwidth]{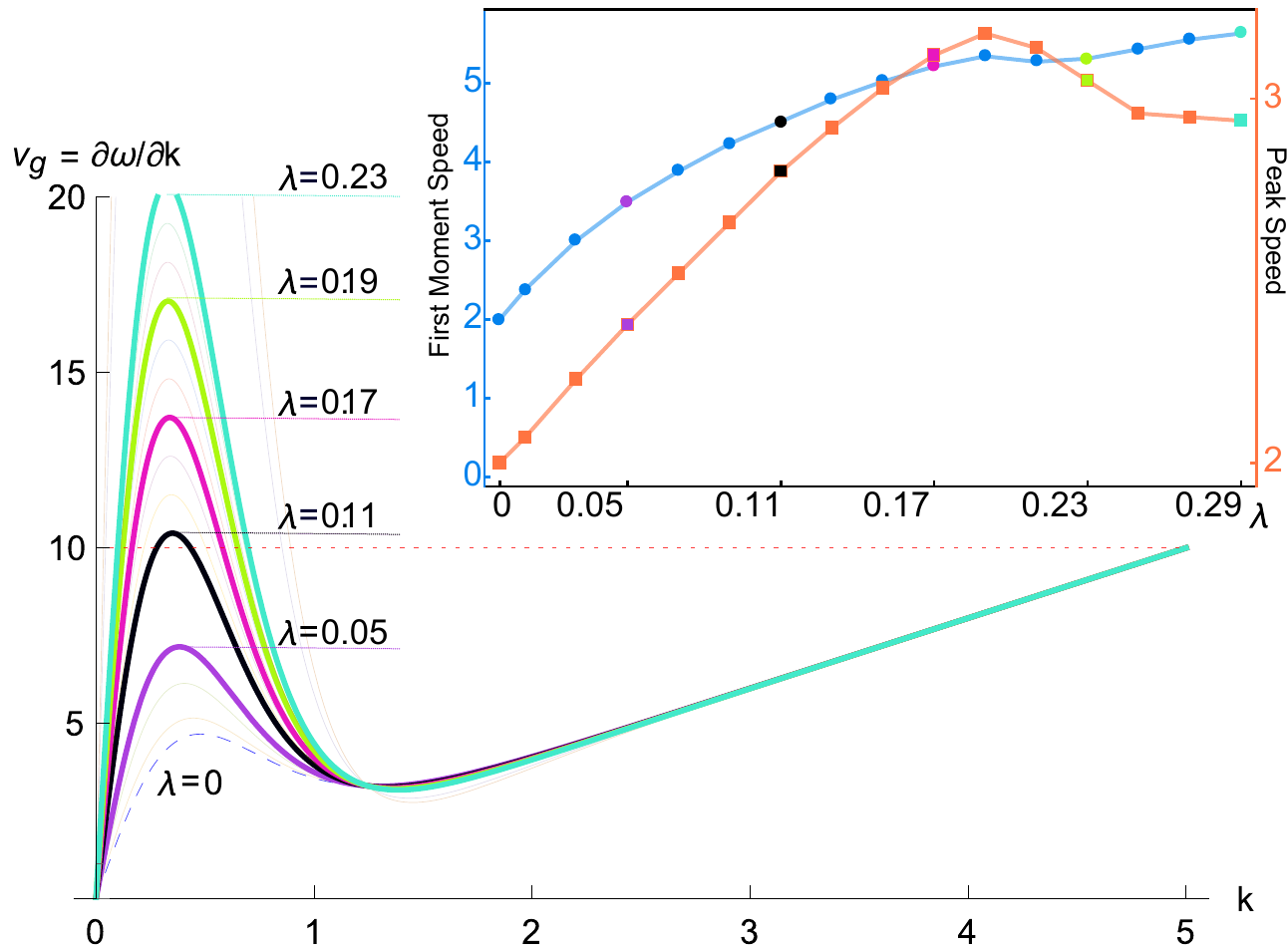}
  \caption{Dispersion of Hasimoto's Soliton. The non-Hamiltonian evolution enhances to group velocity of the long wavelength modes in an initial soliton state. The speed of the curvature peak and first moment (inset) are increasing functions of $\lambda$ until roughly $\lambda=0.19$ where a majority of shape defining Fourier modes leave an eroding peak to bolster the first moment. 
  }
  \label{fig:dispersion}
\end{figure}

A plane wave solution of the form $\psi = A e^{i(k s- \omega t)}$ defines a single mode helix. According to the NVC plane waves obey the corrected nonlinear dispersion relation~[\onlinecite{Newton1987StabilityWaves}],
 \begin{align}\label{eqn:Dispersion}
  \omega(k,A,\lambda) = k^{2}(1+ \lambda A^{2}) - \frac{A^{2}}{2} - \frac{3 \lambda}{4} A^{4}. 
 \end{align}
The initial state of the Hasimoto soliton is given by $\psi = 2 \mbox{sech}(s) e^{is}$, which defines a narrow-band curvature packet, with over  $99\%$ of its total {initial} bending captured between wave numbers $k\in[0,5]$.  Relating the wave amplitude, $A$, to wavenumber via Fourier transform allows us to plot the group velocity for the initial data, which is given in Fig. (\ref{fig:dispersion}). These data show that increasing $\lambda$ enhances the  propagation speed of long wavelength curvature modes. Enhancing dispersion of these modes causes the curvature function to distort. {Simulations indicate that the peak jettisons curvature, which causes the first moment of the distribution to propagate faster, distorting the distribution into a log-normal form.} Additionally, the simulations depict a localized curvature peak that stays discernible under considerable dispersion because of the support provided by the emergent gain/loss mechanism. If total bending were conserved, then the dynamic would cause the peak amplitude to erode completely into the vortex. 

Our non-Hamiltonian vortex dynamics can be simulated through the vector evolution equation, Eq.~(\ref{eqn:BNF}), on a mesh of points representing the vortex in $\mathbb{R}^{3}$, or by scalar evolution of the geometric variables through the NVC. In this way the Hasimoto transformation effectively separates the geometric evolution from its Frenet-Serret representation and is significantly more efficient if vortex visualization is not required. In addition to the heightened regularity requirements on the geometric variables for the Hasimoto transformation, it should be noted that the boundary conditions for each problem are physically different. Specifically, for an arc-length variable $s\in[a,b]$,  Dirichlet boundary conditions, $\psi(a,t)=\psi(b,t)=0$, allow for freely moving endpoints with zero curvature.  In contrast, Dirichlet conditions on the vector evolution fix the endpoints, $\vec{\gamma}(a,t)=\vec{\gamma}_{a}$ and $\vec{\gamma}(b,t)=\vec{\gamma}_{b}$ where $\vec{\gamma}_{a},\vec{\gamma}_{b}\in\mathbb{R}^{3}$.  To suppresses any differences manifesting from the endpoint behavior, we simulate Hasimoto's soliton under both the vector and scalar evolutions on an arc-length domain an order of magnitude larger than the characteristic width of this curvature disturbance~\bibnote{
Our simulations utilize the \texttt{NDSolve} routine of Mathematica, set to work at 10 digits of internal precision and interpolating at the order of the underlying method chosen to  numerically integrate the system of ordinary differential equations arrived at by application of the method of lines.
}. {The total bending was calculated for both methods and found to have less than 1\% squared relative error in the curvature across the lifetime of the simulation.}  Our focus is on simulations where $\lambda = 0.17$, which displays features types of non-Hamiltonian waves, {and corresponds to flows generated by arcs whose length is near the regularization limit.} 

\begin{figure}
\includegraphics[width=.5\textwidth]{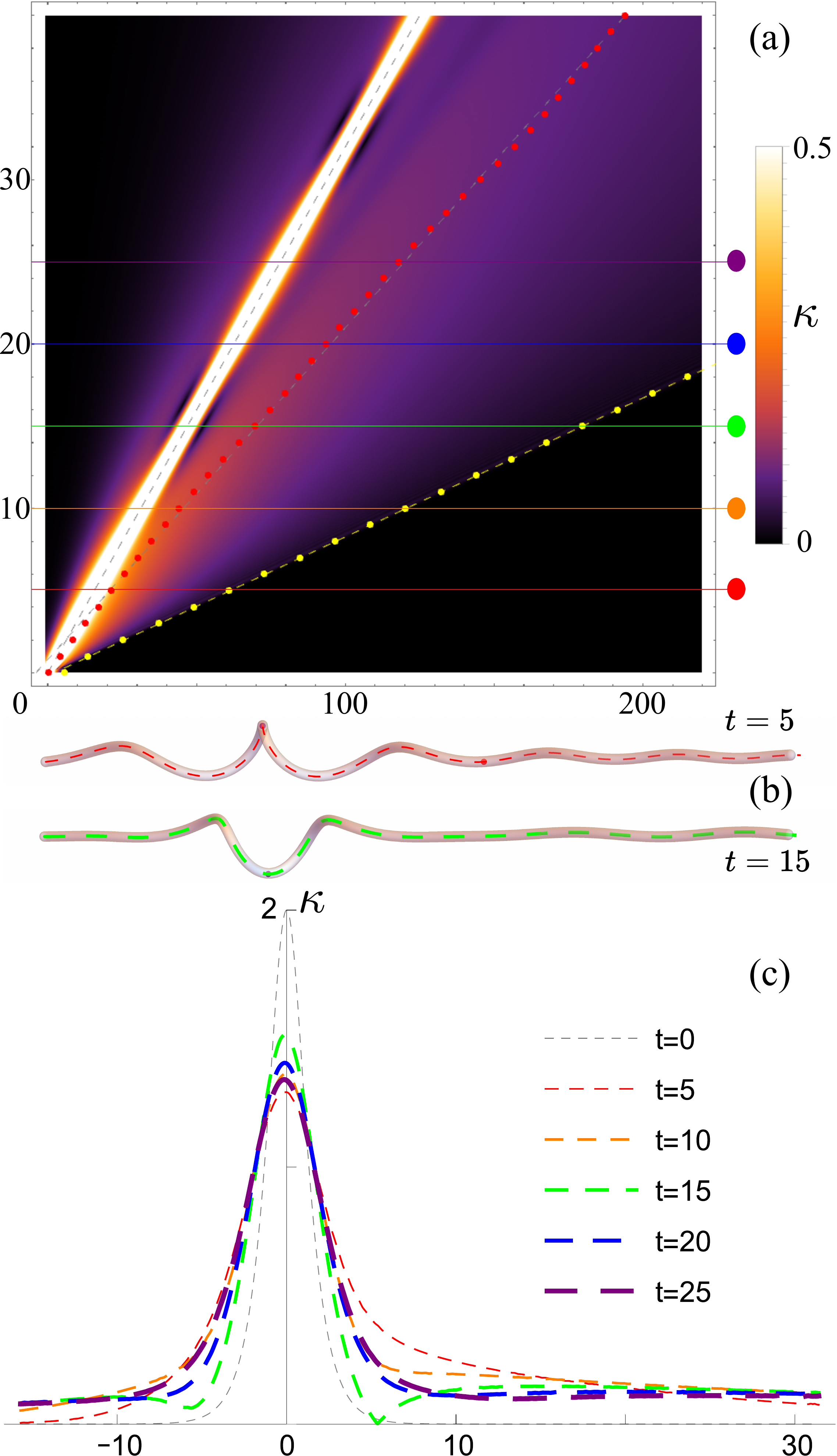}
\caption{Cascade Dynamics. (a) For moderate dispersion the cascade soliton has a well defined peak as it breaths and disperses.  The clearly defined peak is tracked along with the first moment (red) and wave front (yellow) on a density plot of curvature. (b) Additionally, we depict vortex configurations at $t=5$ (orange), $t=15$ (green) which illustrates the breathing dynamic. (c) Lastly, a sequence of curvatures at five-second intervals are plotted against the initial state, adjusted for translations, and show a clear asymmetry in the profile in addition to the breathing.
}
\label{fig:17Density}
\end{figure}

{The NVC predicts} two qualitatively different dynamical regimes, one characterized {the cascade soliton} and the other by the strong dispersion of this dissipative soliton. Figure (\ref{fig:17Density}) plots curvature as a density for $\lambda=0.17$  and we see that the maximum amplitude is strongly localized while the bending energy disperses in the direction of peak propagation. In Fig.~(\ref{fig:17Density}a) we plot the maximum value, first moment (red) and wave front (yellow), i.e., the furthest point ahead of the peak were $\kappa\approx 2\%$ of the initial peak value are fitted to linear models with square residuals, $0.99999$, $0.999198$, $0.998574$, respectively. Our simulations also show the presence of breathing oscillations, seen by the dark side bands to the peak appearing twice. In Fig.~(\ref{fig:17Density}c) the middle of a breath can be seen as pinch in the curvature function occurring at $t=15$ and is plotted along with curvatures at $t=0,5,10,25$  in the co--moving frame. Simulations omitting the $\left[|\psi|^{2} \psi\right]_{ss}$ term exhibit a similar  dynamic and  indicates that breathing is, in part, a consequence of the quintic nonlinearity. The completion of two  breaths was corroborated with a power--spectrum analysis of the time data. In addition to this breathing, the gain/loss mechanism creates an asymmetry in the curvature profile that when coupled to dispersion leads to a trailing helical wake of low wavenumber curvature modes. The dissipative{/cascade} soliton corresponds to the peak amplitude following the helical excitations. Additional simulations show that for $\lambda < 0.17$ we see a similar dynamic, but the peak is strongly maintained and less curvature is dispersed. Also, for $\lambda > 0.17$ the peak speed begins to decrease as low wavenumber modes shift the curvature distribution to a log-normal form. While the curvature peak is discernible, it is difficult to spot immersed in a sea of Kelvin waves. Additionally, the breathing is abated on the simulated time scale of $40$ seconds. {If a condensate is punctured by a vortex defect with circulation $997\times 10^{-4} \textrm{$\mu$m} \cdot \textrm{cm}$, length $100\textrm{$\mu$m}$ and core size of $0.67\textrm{$\mu$m}$, then for an vortex element whose radius of curvature is $12\textrm{$\mu$m}$, a reasonable size for a vortex ring,  Eq.~(\ref{eqn:SV}) yields a characteristic time scale on the order of milliseconds, which is with the range of times considered in reconnection studies~[\onlinecite{Kivotides2001KelvinTurbulence}-\onlinecite{Paoletti2010ReconnectionVortices}].}

Understanding that the Biot-Savart integral is a manifestation of the mean-field Gross-Pitaevskii dynamics of the condensate, our beyond local induction model self-consistently describes the dynamics of isolated quantized vortices whose flow is induced at length scales nearing the healing length. While this non-integrable and computationally inexpensive result can be easily added to current filament models, it is also a useful symbolic tool for investigation of post reconnection dynamics~[\onlinecite{Hanninen2014VortexTurbulence.}]. One expects that such events are the driving mechanism of highly localized curvature distributions which are now experimentally realizable for condensates with a few vortex defects~[\onlinecite{Serafini2017VortexCondensates}]. Perhaps through minimally defected flows we can gain greater insight into Onsager's conjectured mechanism of anomalous dissipation which asserts that weak solutions of inviscid fluid dynamics are not necessarily conservative and that the geometry itself is capable of relaxation by radiating turbulent energy toward the fluid boundaries~[\onlinecite{Eyink2006OnsagerTurbulence}]. 

In conclusion, we derive a non-Hamiltonian evolution for the curvature and torsion of a quantized vortex that breaks the integrability of the local induction approximation and introduces a helical shock wave on the vortex medium. Such a dynamic is necessary to support the cascade process associated with low temperature quantum turbulence. The shock consists of a leading packet of Kelvin waves dispersed from a dissipative vortex soliton, i.e., a non-Hamiltonian cascade soliton. 

The authors acknowledge support from the US National Science Foundation under grant numbers PHY-1306638, PHY-1207881, PHY-1520915, OAC-1740130, and the US Air Force Office of Scientific Research grant number FA9550-14-1-0287. This work was performed in part at the Aspen Center for Physics, which is supported by the US National Science Foundation grant PHY-1607611.

\end{document}